# Comment on "Fractional quantum mechanics" and "Fractional Schrödinger equation"

Yuchuan Wei[*]

Nanmengqian, Dafeng, Wuzhi, Jiaozuo, Henan 454981, China

*Corresponding author: yuchuanwei@gmail.com

In this comment, we point out some shortcomings in two papers "Fractional quantum mechanics" [Phys. Rev. E 62, 3135 (2000)] and "Fractional Schrödinger equation" [Phys. Rev. E 66, 056108 (2002)]. We prove that the fractional uncertainty relation does not hold generally. The probability continuity equation in fractional quantum mechanics has a missing source term, which leads to particle teleportation, i.e., a particle can teleport from one place to another. Since the relativistic kinetic energy can be viewed as an approximate realization of the fractional kinetic energy, the particle teleportation should be an observable relativistic effect in quantum mechanics. With the help of this concept, superconductivity could be viewed as the teleportation of electrons from one side of a superconductor to another and superfluidity could be viewed as the teleportation of helium atoms from one end of a capillary tube to the other. We also point out how to teleport a particle to a destination.

## I. Introduction

Historically quantum mechanics based on general kinetic energy has been studied widely [1]. In Refs [2,3], the standard quantum mechanics [4] was generalized to fractional quantum mechanics. The Schrödinger equation was rewritten as

$$i\hbar \frac{\partial}{\partial t} \psi(\mathbf{r},t) = H_\alpha \psi(\mathbf{r},t)$$
$$H_\alpha = T_\alpha + V = D_\alpha |\mathbf{p}|^\alpha + V(\mathbf{r}).$$
(1)

As usual, $\psi(\mathbf{r},t)$ is a wave function defined in the 3 dimensional space and dependent on time $t$, $D_\alpha$ is a constant dependent on the fractional parameter $1 < \alpha \leq 2$, $\mathbf{r}, \mathbf{p}$ are the position and momentum operators, respectively, $\hbar$ is the Plank constant, and m is the mass of a particle. The fractional Hamiltonian operator $H_\alpha$ is the sum of the fractional kinetic energy $T_\alpha$ and the potential energy $V(\mathbf{r})$.

When $\alpha = 2$, taking $D_2 = 1/(2m)$, the fractional kinetic energy becomes the classical kinetic energy

$$T_2 = \frac{\mathbf{p}^2}{2m} = T \qquad (2)$$

and the fractional Schrödinger equation becomes the standard Schrödinger equation. When $1 < \alpha < 2$, the fractional kinetic energy operator is defined by the momentum representation [2].

However, there exist three shortcomings in this recent quantum theory:

(1) The Heisenberg uncertainty relation was generalized to the fractional uncertainty relation [2]

$$<|\Delta x|^\mu>^{1/\mu} <|\Delta p|^\mu>^{1/\mu} \; > \; \frac{\hbar}{(2\alpha)^{1/\mu}}, \quad \mu<\alpha, 1<\alpha \leq 2. \qquad (3)$$

It seems unsuitable to call this inequality fractional uncertainty relation, and this inequality does not hold mathematically.

(2) The fractional probability continuity equation obtained by Laskin [3] was

$$\frac{\partial}{\partial t}\rho + \nabla \cdot \mathbf{j}_\alpha = 0 \qquad (4)$$

where the probability density and current density were defined as

$$\begin{aligned}\rho &= \psi^*\psi \\ \mathbf{j}_\alpha &= -iD_\alpha \hbar^{\alpha-1}\left(\psi^*(-\nabla^2)^{\alpha/2-1}\nabla\psi - \psi(-\nabla^2)^{\alpha/2-1}\nabla\psi^*\right).\end{aligned} \qquad (5)$$

In fact, a source term was missing, which indicates another way of probability transportation, probability teleportation.

(3) The relationship between fractional quantum mechanics and the real world was not given, and it was almost impossible to find the applications of this theory. Here we will point out that the relativistic kinetic energy can be viewed as an approximate realization of the fractional kinetic energy, which makes the probability teleportation a practical phenomenon.

Now we will discuss these shortcomings in order. For the convenience, please be reminded that the symbol $H^+$ in [2,3] should be H.

## II. Fractional Uncertainty Relation

### 1. The uncertainty relation is independent of wave equations

For simplicity, we do not consider wave functions which are not square integrable.

Suppose that $\psi(x)$ is a normalized square integrable wavefunction defined on the x-axis. Heisenberg's uncertainty relation says [4]

$$\sqrt{<(\Delta x)^2>} \sqrt{<(\Delta p)^2>} \geq \frac{\hbar}{2}, \tag{6}$$

where

$$\Delta x = x - <x>, \quad \Delta p = p - <p>. \tag{7}$$

As usual, $x$, $p$ stand for the 1D position and momentum operator, and $<x>$, $<p>$ stand for their averages on the wave function $\psi(x)$, for example,

$$<p> = \int_{-\infty}^{\infty} \psi^*(x)\left(-i\hbar \frac{\partial}{\partial x}\right)\psi(x)dx. \tag{8}$$

This relation holds for all the square integral functions, and it is a property of the space of square-integrable functions. A complete mathematical proof can be seen in [5].

As a kinetic equation, the Schrödinger equation

$$i\hbar \frac{\partial}{\partial t}\psi(x,t) = H\psi(x,t) \tag{9}$$

tells us how to determine the wavefunction-time relation $\psi(x,t)$ by the Hamiltonian operator $H$ given the initial wave function $\psi(x,0)$. From the viewpoint of geometry, the equation (9) defines a curve in the space of square integrable functions, which passes a given point at time t=0. Heisenberg's uncertainty relation and the Schrödinger equation are independent.

Laskin generalized the Schrödinger equation, but wave functions remains square-integrable functions, in other words the used function space remains the space of the square-integrable functions, so the Heisenberg uncertainty relation remains true, regardless of the standard or fractional quantum mechanics.

In addition, suppose that there is an uncertainty relation, which holds for all the solutions of the fractional Schrödinger equation (1) with certain $\alpha$, e.g. $\alpha = 1.5$. Since the initial wavefunction $\psi(x,0)$ is an arbitrary square integrable function, we know that this uncertainty relation holds for the whole space of square-integrable functions. Therefore there does not exist a so-called fractional uncertainty relation.

Generally speaking, a generalization [1] of the Schrödinger equation does not generate new uncertainty relations if the wavefunctions remains square-integrable functions.

## 2. Fractional uncertainty relation does not hold in mathematics

Even with the Levy wave packet (Equation 35 in [2]), the uncertainty relation (3) does not hold in the sense of mathematics. We prove it by contradiction. There are two steps.

(1) Let us consider the case $\mu = 1$ and $\alpha = 1$ first.

The Levy wave packet with $\nu = 1$ at $t=0$ is

$$\psi_L(x,0) = \frac{1}{2\hbar}\sqrt{\frac{l}{\pi}}\int_{-\infty}^{\infty}\exp(-\frac{|p-p_0|l}{2\hbar})\exp(i\frac{p}{\hbar}x)dp$$
$$= \frac{1}{2}\sqrt{\frac{l^3}{\pi}}\frac{1}{x^2+(l/2)^2}\exp(i\frac{p_0}{\hbar}x).$$
(10)

The letters $L$ means the Levy wave packet and $l$ is a reference length.

The related quantities can be calculated as

$$<x> = 0, \quad <p> = p_0,$$
$$<|\Delta x|> = <|x|> = \int_{-\infty}^{\infty}|x|\psi_L^*(x,0)\psi_L(x,0)dx = \frac{l}{\pi},$$
$$<|\Delta p|> = <|p-p_0|> = \frac{l}{2\hbar}\int_{-\infty}^{\infty}|p-p_0|\exp(-\frac{|p-p_0|l}{\hbar})dp = \frac{\hbar}{l}.$$
(11)

Therefore we have the inequality

$$<|\Delta x|> \; <|\Delta p|> \; = \; \frac{l}{\pi}\frac{\hbar}{l} = \frac{\hbar}{\pi} < \frac{\hbar}{2}.$$
(12)

(2) At $t=0$, keep $\mu = 1$ as a constant and let $\alpha \to 1^+$.

Since the parameter of the Levy wave packet $\nu = \alpha$, the two sides of the inequality (3) are continuous functions about $\alpha$. Taking $\lim_{\alpha \to 1^+}$ of the both sides of the fractional uncertainty relation (3), we get

$$<|\Delta x|> \; <|\Delta p|> \; \geq \; \frac{\hbar}{2}$$
(13)

which contradicts inequality (12). Therefore the fractional uncertainty relation (3) does not hold mathematically.

Further, once the fractional uncertainty relation does not hold for certain $\alpha$ at $t=0$, we know that there exists a small time neighborhood [0,δ), the relation does not hold either, since the wave packet has not expanded very much. In short, the fractional generalization of the Heisenberg uncertainty relation does not hold generally.

We would like to explain why we could take $\alpha = 1$, which was not included in [2,3]. The case $\alpha = 1$ is just a step of our proof, like an auxiliary line used in geometry problems. Here we add two points. 1) There exist papers that allow $0 < \alpha \leq 2$. In [6], Jeng *et al* claimed that Laskin's solutions for the infinite square well problem were wrong by means of the evidence from the case $0 < \alpha < 1$. In fact the evidence from the case $\alpha = 1$ is more straightforward [7]. (2) The fractional Schrödinger equation with $\alpha = 1$ has many closed-form solutions [8], which is an easy starting point for the study of the fractional Schrödinger equation with $1 < \alpha < 2$.

## III. Probability Continuity Equation

In this section, we will present the correct probability continuity equations in the fractional quantum mechanics, and reveal a different phenomenon of the probability transportation.

**1. The correct probability continuity equation**

From the fraction Schrödinger equation (1), we can get

$$i\hbar \frac{\partial}{\partial t}(\psi^*\psi) = \psi^* T_\alpha \psi - \psi T_\alpha \psi^*. \tag{14}$$

According to Laskin's definitions of the probability density and the current density(5), the correct probability continuity equation

$$\frac{\partial}{\partial t}\rho + \nabla \cdot \mathbf{j}_\alpha = I_\alpha \tag{15}$$

has an extra source term

$$I_\alpha = -iD_\alpha \hbar^{\alpha-1}\left(\nabla \psi^*(-\nabla^2)^{\alpha/2-1}\nabla \psi - \nabla \psi(-\nabla^2)^{\alpha/2-1}\nabla \psi^*\right). \tag{16}$$

Specifically, if $I_\alpha(\mathbf{r},t) > 0$, there is a source at position **r** and time t, which generates the probability; when $I_\alpha(\mathbf{r},t) < 0$, there is a sink at position **r** and time t, which destroys the probability.

It is easy to find cases where the source term is not zero. For example, take the wave function

$$\begin{aligned}\psi &= \psi_1 + \psi_2 \\ \psi_1(x,t) &= \exp(ik_1 x)\exp(-iE_1 t) \\ \psi_2(x,t) &= \exp(ik_2 x)\exp(-iE_2 t)\end{aligned} \tag{17}$$

with $k_1 > k_2 > 0, E_1 = D_\alpha(\hbar k_1)^\alpha, E_2 = D_\alpha(\hbar k_2)^\alpha$, which is a superposition of two solutions to the fractional Schrödinger equation for a free particle.

We have

$$\begin{aligned}
I_\alpha &= -iD_\alpha \hbar^{\alpha-1} \left( \nabla \psi^* (-\nabla^2)^{\alpha/2-1} \nabla \psi - \nabla \psi (-\nabla^2)^{\alpha/2-1} \nabla \psi^* \right) \\
&= -iD_\alpha \hbar^{\alpha-1} \left[ \left( \psi_1^* + \psi_2^* \right)' (-\nabla^2)^{\alpha/2-1} \left( \psi_1 + \psi_2 \right)' - \left( \psi_1 + \psi_2 \right)' (-\nabla^2)^{\alpha/2-1} \left( \psi_1^* + \psi_2^* \right)' \right] \\
&= -iD_\alpha \hbar^{\alpha-1} \left( \psi_1^{*'} (-\nabla^2)^{\alpha/2-1} \psi_2' + \psi_2^{*'} (-\nabla^2)^{\alpha/2-1} \psi_1' - \psi_1' (-\nabla^2)^{\alpha/2-1} \psi_2^{*'} - \psi_2' (-\nabla^2)^{\alpha/2-1} \psi_1^{*'} \right) \quad (18) \\
&= -iD_\alpha \hbar^{\alpha-1} \left( k_1 k_2^{\alpha-1} \psi_1^* \psi_2 + k_1^{\alpha-1} k_2 \psi_2^* \psi_1 - k_1 k_2^{\alpha-1} \psi_1 \psi_2^* - k_1^{\alpha-1} k_2 \psi_2 \psi_1^* \right) \\
&= -iD_\alpha \hbar^{\alpha-1} (k_1 k_2^{\alpha-1} - k_1^{\alpha-1} k_2)(\psi_1^* \psi_2 - \psi_1 \psi_2^*) \\
&= 2D_\alpha \hbar^{\alpha-1} (k_1 k_2^{\alpha-1} - k_1^{\alpha-1} k_2) \sin\left( (k_2 - k_1)x - (E_2 - E_1)t/\hbar \right),
\end{aligned}$$

which is not zero unless $\alpha = 2$.

The source term indicates that the probability is no longer locally conserved. As Laskin proved in [9], the total probability in the whole space is conserved. Here the probability transportation in the fractional quantum mechanics becomes unusual: Some probabilities can disappear at a region and simultaneously appear at other regions but the total probability does not change. In other words, some probabilities can teleport from one place to another. Furthermore, if the particle has mass and charge, probability teleportation will imply mass teleportation and charge teleportation. We need to pay close attention to this phenomenon, as mass teleportation contradicts our life experience and charge teleportation contradicts the classical electrodynamics.

**2. The case $I_\alpha(\mathbf{r}, t) = 0$**

When $\alpha = 2$, it is easy to see $I_2(\mathbf{r}, t) = 0$. The fractional continuity equation recovers the standard continuity equation.

**Proposition.** For $1 < \alpha < 2$, we have $I_\alpha(\mathbf{r}, t) = 0$ for a free particle with a definite kinetic energy.

Proof.

Since $V(\mathbf{r}) = 0$, the fractional Schrödinger equation is

$$i\hbar \frac{\partial}{\partial t} \psi(\mathbf{r}, t) = T_\alpha \psi(\mathbf{r}, t). \quad (19)$$

For a definite energy E, its solution is

$$\psi(\mathbf{r}, t) = \int_\Omega C(\theta, \phi) \exp(i\mathbf{k} \cdot \mathbf{r}) \sin\theta \, d\theta \, d\phi \exp(-iEt/\hbar), \quad (20)$$

$$E = D_\alpha (\hbar k)^\alpha. \tag{21}$$

where $\Omega$ is the unit sphere, $(k,\theta,\phi)$ is the spherical coordinate of the wave vector $\mathbf{k}$, and $C(\theta,\phi)$ is an arbitrary function.

Thus we have

$$(-\nabla^2)^{\alpha/2-1}\psi(\mathbf{r},t) = k^{\alpha-2}\psi(\mathbf{r},t) \tag{22}$$

$$(-\nabla^2)^{\alpha/2-1}\psi^*(\mathbf{r},t) = k^{\alpha-2}\psi^*(\mathbf{r},t). \tag{23}$$

In this case the source term vanishes

$$\begin{aligned}
I_\alpha &= -iD_\alpha \hbar^{\alpha-1}\left(\nabla\psi^*(-\nabla^2)^{\alpha/2-1}\nabla\psi - \nabla\psi(-\nabla^2)^{\alpha/2-1}\nabla\psi^*\right) \\
&= -iD_\alpha \hbar^{\alpha-1}\left(\nabla\psi^*\nabla(-\nabla^2)^{\alpha/2-1}\psi - \nabla\psi\nabla(-\nabla^2)^{\alpha/2-1}\psi^*\right) \\
&= -iD_\alpha \hbar^{\alpha-1}k^{\alpha-2}\left(\nabla\psi^*\nabla\psi - \nabla\psi\nabla\psi^*\right) = 0,
\end{aligned} \tag{24}$$

and the continuity equation has a sourceless form

$$\frac{\partial}{\partial t}\rho + \nabla \cdot \mathbf{j}_\alpha = 0, \tag{25}$$

with

$$\begin{aligned}
\rho &= \psi^*\psi \\
\mathbf{j}_\alpha &= -iD_\alpha \hbar^{\alpha-1}k^{\alpha-2}\left(\psi^*\nabla\psi - \psi\nabla\psi^*\right).
\end{aligned} \tag{26}$$

This completes the proof.

We emphasize that in scatter problems the source term $I_{1<\alpha<2}(\mathbf{r},t) \neq 0$ at the detector's location, though the potential at the detector may be zero. There are two reasons: 1) The particle is not free so the relation (22) does not hold, and 2) the kinetic energy of particles from the scattering source may not be exactly the same, i.e. they may not be strictly monoenergetic.

How to develop a scattering model based on the correct continuity equation (15) is a valuable problem in quantum mechanics.

### IV. Schrödinger Equation with Relativistic Kinetic Energy

In Refs [2,3], the relationship between the fractional quantum mechanics and the real world was not given. A natural question is which particle has a fractional kinetic energy. If there are no fractional particles in our world, why do we need fractional quantum mechanics? To relate the fractional quantum mechanics to the real world, we regard the relativistic quantum mechanics [1,10-12] as an approximate realization of fractional quantum mechanics.

According to the special relativity, the relativistic kinetic energy is

$$T_r = \sqrt{\mathbf{p}^2 c^2 + m^2 c^4}, \tag{27}$$

where the subscript r means special relativity.

For the case of low speed, the relativistic kinetic energy is approximately the summation of the rest energy and the classical kinetic energy ($\alpha = 2$)

$$T_r \approx mc^2 + \frac{\mathbf{p}^2}{2m} = mc^2 + T_2, \tag{28}$$

and for the case of extremely high speed, where the rest energy can be neglected, the relativistic kinetic energy is the fractional kinetic energy with $\alpha = 1$

$$T_r \approx |\mathbf{p}|c = T_1. \tag{29}$$

Generally speaking, if the speed of a particle increases from low to high, the relativistic kinetic energy $T_r$ will approximately correspond to a fractional kinetic energy $T_\alpha$, whose parameter $\alpha$ changes from 2 to 1. Therefore the relativistic kinetic energy is an approximate realization of the fractional kinetic energy.

The Hamiltonian function with the relativistic kinetic energy is [10,11]

$$H_r = \sqrt{\mathbf{p}^2 c^2 + m^2 c^4} + V(\mathbf{r}). \tag{30}$$

Historically, using this Hamiltonian function, Sommerfeld calculated the relativistic correction to Bohr's hydrogen energy levels, and the fine structure in the hydrogen spectrum was explained exactly [4].

The relativistic Schrödinger equation is [1, 10,11]

$$i\hbar \frac{\partial}{\partial t} \psi(\mathbf{r},t) = H_r \psi(\mathbf{r},t). \tag{31}$$

Using the perturbation method, we recently calculated the relativistic correction to the hydrogen energy levels obtained from the Schrödinger equation. The resultant energy levels contain an $\alpha^5$ term, which can explain the Lamb shift at an accuracy of 41% [13,14].

The continuity equation can be expressed as [1]

$$\frac{\partial}{\partial t}\rho + \nabla \cdot \mathbf{j}_r = I_r,  \tag{32}$$

with the current density and the source term

$$\begin{aligned}\mathbf{j}_r &= -\frac{1}{i\hbar}(\psi^* T_r \nabla^{-2}\nabla \psi - \psi T_r \nabla^{-2}\nabla \psi^*), \\ I_r &= -\frac{1}{i\hbar}(\nabla \psi^* T_\alpha \nabla^{-2}\nabla \psi - \nabla \psi T_\alpha \nabla^{-2}\nabla \psi^*).\end{aligned} \tag{33}$$

Again, the probability is not locally conserved, but the total probability in the whole space is conserved [1]

$$i\hbar \frac{\partial}{\partial t}\int_{R^3}\psi^*\psi d^3\mathbf{r} = \int_{R^3}\psi^* T_r \psi d^3\mathbf{r} - \int_{R^3}\psi T_r \psi^* d^3\mathbf{r} = 0. \tag{34}$$

Similarly, for a free particle with a definite kinetic energy, we have $I_r = 0$.

Since the relativistic kinetic energy is true and the classical kinetic energy is approximate, the probability continuity equation with the source term (32) can be true and the popular probability continuity equation in standard quantum mechanics [4] is approximate.

Therefore we need to base our scattering model on the continuity equation with the source term, i.e. Eq. (32), calculate the variation between the present model and the traditional model, and design experiments to observe the phenomenon of the probability teleportation.

Since the relativistic Schrödinger equation (31) is not relativistically covariant [10,11], violates the causality [15], and is non-local [16] and complicated [10], the research on this equation has been criticized since the early days of quantum mechanics. A positive experimental result on the probability teleportation will end this situation ultimately.

## VI. Conclusion

We prove that the fractional uncertainty relation does not hold generally. The missing source term in the fractional probability continuity equation leads to particle teleportation. According to the special relativity, the classical kinetic energy is just an approximation in the low speed case, so it would be a hopeful direction to study particle teleportation in scattering theory and experiments. Furthermore, the concept in this comment offers a very intuitive explanation for superconductivity and superfluidity. We also point out how to teleport a particle to a destination.

**Acknowledgement**.

I would like to thank Professor Laskin for alerting me to Ref. [17]. I would like to thank the reviewers for their suggestions on revision. Also I would like to thank Prof. Xiaofeng Yang and Wengfeng Chen for the discussion on the relativistic covariance.

## Appendix A. An intuitive explanation for the superfluidity and superconductivity based on teleportation.

In [17] the authors applied the fractional quantum mechanics to explain some property of the superfluid $^4$He.

From fractional Schrödinger equation, they correctly got the fractional probability continuity equation

$$\frac{\partial}{\partial t}\rho + \nabla \cdot \mathbf{j}_\alpha = I_\alpha, \qquad (35)$$

where $\rho$ was viewed as the density of the superfluid, $\mathbf{j}_\alpha$ as the mass current density, and $I_\alpha$ as extra sources.

In order to keep consistent with the well-known fluid continuity equation

$$\frac{\partial}{\partial t}\rho + \nabla \cdot \mathbf{j} = 0, \qquad (36)$$

the author claimed that in the superfluid $I_\alpha \approx 0$, since the wavefunction describing the He atoms could be assumed to be

$$\psi(\mathbf{r}) = \sum_{\mathbf{p}} C_{\mathbf{p}} \exp(i\mathbf{p}\cdot\mathbf{r}/\hbar), \qquad (37)$$

where the summation goes over the momentums with approximately equal $|\mathbf{p}|$.

This is very difficult to understand. First, we do not know why the atoms are on such a special state. Second, since the potential $V(\mathbf{r})$ is not zero or a constant, (see Eq. (13) in [17]), the wave function (37) is not an eigenfunction of the fractional Hamiltonian operator $H_\alpha$, so even if at t=0 the momentum magnitudes $|\mathbf{p}|$ are the same, they will become different soon. On the contrary, we should suppose that the source term $I_\alpha$ or $I_r$ is not zero generally.

Here is our superfluid model based on equation (35) or (32). When the superfluid $^4$He is still, moving atoms can disappear at one place and appear at another distant place. This explains why heat can be conducted easily by the superfluid. When the superfluid flows, some superfluid moves forward normally, which has friction, and some teleports from one place to another place, which has no friction. Thus we do not need to artificially divide the superfluid into a normal component and a superfluid component.

Now it becomes an urgent task to observe whether the mass teleportation really exists in the superfluid experiment. An easy way is to measure the velocity of the superfluid in the capillary tube and the mass coming out from the tube to check whether they are consistent.

Similarly, based on the concept of teleportation, the superconductivity can be viewed as such a phenomenon in which some electrons teleport from one side of the superconductor to the other side; of course they do not dissipate energy. Currently, physicists say that the superconducting electrons pass through the Josephson junction by quantum tunneling, but an open question is why the supercurrent can flow through the non-superconducting metal in an SNS (Superconductor-NormalMetal-Superconductor) junction **without dissipation**. (How can one drive through a toll tunnel without payment?) The reasonable explanation should be that electrons teleport from one side of the junction to the other.

**Appendix B. Particle teleportation based on wave function collapse**

In current quantum teleportation, the state of a particle is transferred from one place to another while the particle itself does not move at all. Here we provide a procedure for particle teleportation based on wave function collapse in quantum mechanics.

Suppose that a particle is located in an interval on the negative half of $x$-axis, described by a wave function

$$\psi(x) = \begin{cases} \text{nonzero} & x \in [-b, -a] \\ 0 & x \notin [-b, -a] \end{cases}.$$

Here a<b are two positive real numbers. Obviously, this wave function is neither even nor odd. Technically, the function is not an eigenfunction of the parity operator, but a superposition of two eigenfunctions

$$\psi(x) = (\psi_{\text{even}}(x) + \psi_{\text{odd}}(x))/2$$

with

$$\psi_{\text{even}}(x) = \psi(x) + \psi(-x), \quad \psi_{\text{odd}}(x) = \psi(x) - \psi(-x).$$

Now we measure the parity of the state, the wave function $\psi(x)$ will collapse into either $\psi_{\text{even}}(x)$ or $\psi_{\text{odd}}(x)$. In either case, the probability of the particle appearing on the positive half of $x$-axis is ½. Now we detect the particle in the region [a, b]. If we find it, the teleportation is over. If we do not find it, we know that the particle remains in the region [-b,-a], and we measure the parity of the wave function $\psi(x)$ and detect the particle in the region [a,b] again. We repeat till we find it. In addition, we point out that the spin state of the particle remains the same after this teleportation.

This procedure reminds us that wave function collapse in quantum mechanics can cause a particle to run faster than light.